%% file: CAS_SR_Paper.tex
\documentclass[a4paper,fleqn]{cas-dc}

\usepackage[numbers]{natbib}

\usepackage{amsmath}
\usepackage{amssymb}
\usepackage{mathtools}
\usepackage{booktabs}
\usepackage{graphicx}
\usepackage{siunitx}
\usepackage{pgfplots}
\usepackage{makecell}
\usepackage{multirow}
\usepackage{tikz}
\usepackage{float}
\usepackage[section]{placeins}         

\usepackage{xcolor}
\input{figs/rwth-colors}

\def\tsc#1{\csdef{#1}{\textsc{\lowercase{#1}}\xspace}}
\tsc{WGM}
\tsc{QE}
\tsc{EP}
\tsc{PMS}
\tsc{BEC}
\tsc{DE}
\usepackage{bookmark}
\hypersetup{
  bookmarksnumbered=true,
  bookmarksopen=true,
  pdfpagemode=UseOutlines
}

\usepackage{subcaption}

\AtBeginDocument{%
  \setlength{\abovedisplayskip}{8pt}%
  \setlength{\belowdisplayskip}{8pt}%
  \setlength{\abovedisplayshortskip}{8pt}%
  \setlength{\belowdisplayshortskip}{8pt}%
}

\begin{document}
\let\WriteBookmarks\relax
\def\floatpagepagefraction{1}
\def\textpagefraction{.001}

\shorttitle{New Closed-Form Approximations in Statistical Mechanics Obtained by Symbolic Regression}
\shortauthors{O. Kunjir et~al.}

\title[mode=title]{New Closed-Form Approximations in Statistical Mechanics Obtained by Symbolic Regression}                      



\author[1]{O. Kunjir}
\cormark[1]
\fnmark[1]
\ead{kunjir@km.rwth-aachen.de}
\credit{Methodology, Software, Investigation, Visualization, Writing -- original draft, Writing -- review \& editing}

\author[1,2]{A. Popova}
\fnmark[1]
\credit{Methodology, Software, Investigation, Visualization, Writing -- original draft, Writing -- review \& editing}

\author[1]{R. Abdusalamov}
\credit{Methodology, Supervision, Writing -- Review and editing}

\author[1,3]{M. Itskov}
\credit{Conceptualization, Writing – Review and editing, Supervision, Project administration}

\affiliation[1]{organization={Department of Continuum Mechanics,
                RWTH Aachen University},
                addressline={Eilfschornsteinstr. 18},
                city={Aachen},
                postcode={52062},
                country={NRW, Germany}}

\affiliation[2]{organization={School of Mathematical and Statistical Sciences, University of Galway},
                addressline={University Road},
                city={Galway},
                country={Ireland}}

\affiliation[3]{organization={School of Computer Science and Mathematics, Keele University},
                city={Keele},
                state={Staffordshire},
                postcode={ST5 5BG},
                country={United Kingdom}}

\cortext[1]{Corresponding authors}
\fntext[1]{These authors contributed equally to this work.}

\begin{abstract}
    Obtaining compact, analytically tractable closed-form expressions for physically complex systems is a fundamental challenge in statistical mechanics, where traditional approximation methods often sacrifice either accuracy or analytical interpretability across the full parameter range of interest. In this work, we address three approximation problems in polymer statistical mechanics, demonstrating that data-driven expression discovery offers a systematic pathway from exact numerical solutions to interpretable closed-form expressions suitable for downstream analytical use. To this end, we conduct a systematic comparative study using two complementary symbolic regression frameworks — the Deep Symbolic Optimization (DSO) framework, which employs a recurrent neural network trained by reinforcement learning, and PyOperon, a genetic programming framework with nonlinear least-squares constant optimization.
        
    We first apply both frameworks to (i) the force--extension relationship of the Worm-Like Chain (WLC) model, where DSO achieves a maximum relative error below $0.15\%$ and PyOperon reduces this further to $0.05\%$, outperforming the Marko--Siggia formula ($\sim$17\%) and the Petrosyan approximation by factors of up to $18$. Then, we consider (ii) the logarithmic factorial $\ln(n!)$. Trained directly on analytical data, both frameworks recover the dominant structure of Stirling's formula without any structural prior. Using a residual learning strategy on a known continued-fraction expansion of the Stirling remainder, both frameworks outperform all nine classical formulas considered, from Laplace (1820) to Chen (2016): for $n \in [1,30]$, PyOperon reduces the maximum relative error by a factor of $1.3\times10^{4}$ relative to the most accurate of them, while DSO recovers the next two levels of the exact continued fraction. Finally, we consider (iii) the non-Gaussian radial probability density function of the freely jointed chain (FJC) model, where a physics-informed residual learning strategy yields improvements by a factor of $11.2$ over the approximation reported by Darabi et al. (2023) and $4.6$ over the steepest-descent approximation. For the FJC problem, the discovered corrections provide closed-form approximations to the non-Gaussian radial density without relying on a Gaussian-chain approximation. Across all three applications, the resulting expressions remain analytically tractable. Collectively, these results demonstrate that DSO and PyOperon serve as powerful and complementary tools for discovering accurate, analytically tractable approximations in statistical and continuum mechanics.
    
\end{abstract}

\begin{graphicalabstract}
\includegraphics{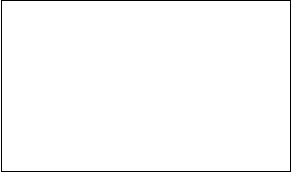}
\end{graphicalabstract}

\begin{highlights}
    \item DSO and PyOperon yield compact closed-form approximations in statistical mechanics
    \item WLC force--extension: maximum error $0.05\%$ with PyOperon, factor of $18$ improvement over Petrosyan
    \item Factorial: PyOperon beats all nine classical formulas by over $10^{4}$ for $n \in [1,30]$
    \item DSO rediscovers the exact continued-fraction structure of the Stirling remainder
    \item FJC density: free PyOperon residual learning is 11.2 times more accurate than Darabi
\end{highlights}


\maketitle

\section{Introduction}
\label{sec:Introduction}
    
    The derivation of compact, closed-form analytical expressions for physically complex models is a fundamental task in both theoretical and applied physics. In statistical mechanics in particular, exact solutions are seldom available, and the advancement of the field has historically relied on the quality of the approximations employed. A suitable approximation must satisfy several competing requirements: it must be sufficiently accurate across the relevant parameter domain, simple enough to be manipulated analytically (integrated, differentiated, inverted), and computationally inexpensive to evaluate. These requirements are frequently in tension, and the search for expressions that reconcile them constitutes a significant body of ongoing research in polymer physics, soft matter science, and related disciplines \cite{Edwards1988,  Rubinstein2003, Treloar2005}.
    
    Symbolic regression (SR) has emerged as a powerful machine learning tool for the data-driven discovery of mathematical expressions. In contrast to classical regression techniques, which fit parameters to a model of pre-assumed functional form, SR searches the combinatorial space of all mathematical expressions constructible from a given library of operators, functions, variables, and constants, yielding an interpretable closed-form equation rather than a black-box model. The field traces its roots to the work of Friedberg \cite{Friedberg1958} and was substantially advanced by the genetic programming (GP) framework of Koza \cite{Koza1992}, which encodes expressions as computational trees subject to evolutionary selection and mutation. Since then, SR has been applied across a wide range of scientific domains, including constitutive modeling \cite{Rasulphd2025, abdusalamov2023automatic, abdusalamov2026rediscovering, abdusalamov2024rediscovering, Bomarito2021, Garbrecht2023}, computational fluid dynamics \cite{angelis2023artificial}, materials science \cite{Ansari2022, burlacu2023symbolic}, and fundamental physics discovery \cite{cranmer2020discovering, la2021contemporary, udrescu2020ai}. However, to the best of our knowledge, application of SR to approximation problems in polymer physics and statistical mechanics appears missing. 
    
    Polymer physics offers a natural and compelling domain for SR-based approximation. Models such as FJC and WLC provide the theoretical foundation for understanding the mechanical behavior of biopolymers including DNA, RNA, actin, and synthetic rubber networks \cite{Bustamante1994, Edwards1988, Treloar2005}. These models are characterized by exact solutions that are either intractable in closed form or require numerical integration, making analytical approximations essential for practical modeling. Critically, errors introduced by existing approximations propagate into resulting physical quantities such as entropy, free energy, and mechanical stress, motivating the development of more accurate yet compact expressions.
    
    This work employs two complementary symbolic regression frameworks. The first is the DSO \cite{dso, petersen2019deep} framework. This framework replaces the classical genetic programming search with a recurrent neural network (RNN) to generate expression trees. With iterative improvement through gradient updates based on the reward signal from evaluated expressions, and guiding the search towards high-reward regions of the expression space it performs more efficiently than evolutionary methods alone. The second is \textit{PyOperon} \cite{burlacu2023symbolic, pyoperon}. It represents Python bindings of the Operon C++ framework, which combines modern genetic programming with nonlinear least-squares constant optimization and efficient parallel execution. Applying both frameworks allows a comparative analysis of their respective strengths across the three problems studied here, and provides cross-validated evidence for the reliability of the discovered expressions.
    
    This paper aims to demonstrate how symbolic regression can be used to derive compact, interpretable analytical approximations for problems in polymer physics and statistical mechanics. Specifically, we obtain a new closed-form approximation of the WLC force--extension relation with the maximum relative error below $0.15\%$. Further, we apply both frameworks to approximate $\ln(n!)$ across real values of $n \in [0, 100]$, improving accuracy for small $n$ over Stirling's formula and benchmarking against nine historical approximations from Laplace (1820) to Chen (2016). Finally, we obtain a compact closed-form approximation of the non-Gaussian radial probability density of the FJC model.
    
    The paper is organized as follows: \autoref{sec:Methodology} introduces the symbolic regression methods, including the DSO and PyOperon frameworks. \autoref{sec:SRFundamentalApproximationProblems} presents the three applications mentioned above. \autoref{sec:Conclusion} summarizes the findings and discusses directions for future work.

\section{Methodology}
\label{sec:Methodology}
    This section introduces the two symbolic regression frameworks employed in this work and describes the implementation details. We begin with a brief overview of symbolic regression and genetic programming, which form the conceptual foundation of both methods. We then describe the DSO \cite{dso, petersen2019deep} and the PyOperon frameworks \cite{burlacu2023symbolic, pyoperon} separately, highlighting the key differences in their search strategies.

\subsection{Symbolic Regression}
\label{sec:SR}
    
    SR is a machine learning technique that seeks a mathematical expression that best describes a relationship between input variables and a target output, without assuming any specific functional form in advance~\cite{la2021contemporary}. Unlike polynomial regression, which restricts the model to a fixed class of functions, or neural network regression, which produces uninterpretable black-box models, SR searches a combinatorial space of expressions composed of elementary operations, for example, addition, multiplication, division, and subtraction, and functions such as exponential, logarithm, trigonometric functions, and power laws. The result is a closed-form algebraic equation, making it directly amenable to analytical manipulation and providing a \emph{white-box} model of the underlying relationship \cite{la2021contemporary}.
    
    The most widely established SR framework is Genetic Programming (GP) \cite{Koza1992}, which maintains a population of candidate expressions and iteratively improves them through four phases: \emph{initialization} (a population of random expressions is generated from a predefined token library), \emph{selection} (expressions are evaluated for fitness and the best individuals are identified via tournament selection), \emph{evolution} (offspring are produced through crossover and mutations), and \emph{termination} (the process repeats for a fixed number of generations or until a fitness threshold is met) \cite{Koza1992, la2021contemporary}.
    
    The evolution phase applies four variation operators to produce offspring expressions: one recombination operator and three mutation operators. \emph{Crossover} swaps a randomly selected subtree from one parent with a subtree from a second donor expression, combining genetic material from both. \emph{Subtree mutation} replaces a randomly selected subtree by a newly generated random subtree, reintroducing diversity without requiring a donor. \emph{Point mutation} replaces a single node by another node of the same arity, introducing minor local variation without altering the global tree structure. \emph{Hoist mutation} replaces a subtree by one of its own child subtrees, shortening the expression and mitigating bloat \cite{Rasulphd2025, Koza1992}. These operators form the core of PyOperon and of the genetic programming component of DSO in its hybrid mode, as described in the following subsections.

\subsection{Deep Symbolic Optimization}
\label{sec:DSO}
    
\begin{figure*}
        \centering
        \includegraphics[width=\textwidth]{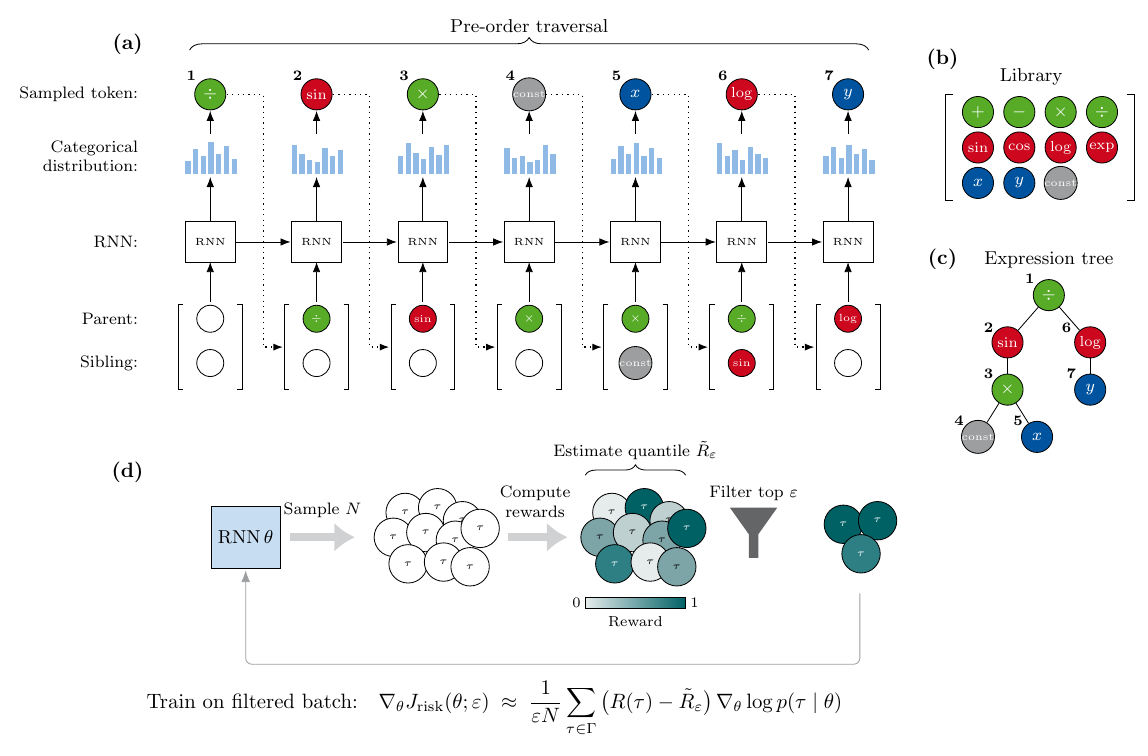}
        \caption{Overview of Deep Symbolic Optimization (DSO).
        \textbf{(a)} Autoregressive generation of an expression by the RNN policy. At each step the RNN emits a categorical distribution over the token library, a token is sampled, and the parent and sibling of the next node (indicated by the dotted arrows) are fed back as the RNN's input, until every branch reaches a terminal. This produces the pre-order traversal $\langle \div,\sin,\times,\text{const},x,\log,y\rangle$. 
        \textbf{(b)} The token library, colored by arity (green = binary operators, red = unary functions, blue = variables, grey = constant).
        \textbf{(c)} The expression tree reconstructed from the traversal, corresponding to $\sin(c\,x)/\log(y)$, where the constant $c$ is fitted to the data.
        \textbf{(d)} The risk-seeking training loop: the RNN samples a batch of $N$ expressions, each receives a reward $R = 1/(1+\mathrm{NRMSE})$, the empirical $(1-\varepsilon)$-quantile $\tilde{R}_\varepsilon$ is estimated, only the top-$\varepsilon$ fraction is kept, and the policy is updated on this filtered batch. Adapted from Petersen et al.~\cite{petersen2019deep} and Mundhenk et al. \cite{mundhenk2021}.}
        \label{fig:dso}
    \end{figure*}

    Deep Symbolic Optimization, proposed by Petersen et al. \cite{dso, mundhenk2021, petersen2019deep}, is a neural-guided symbolic regression framework in which a recurrent neural network (RNN) serves as the primary search mechanism. Unlike pure genetic programming, where the dominant search strategy is evolutionary mutation and crossover, DSO trains the RNN through reinforcement learning (RL) \cite{sutton2018} to generate high-quality expressions directly from a learned probability distribution over the token library. The complete procedure is illustrated in \autoref{fig:dso}.
    
    In the DSO framework, expression trees are encoded as pre-order (prefix) traversals: the root token is sampled first, followed by child tokens from left to right. At each step $t$, the RNN receives an observation consisting of the identity of the parent node and sibling node of the current position in the traversal (\autoref{fig:dso}a). This hierarchical context allows the RNN to respect the structural constraints of valid expression trees (e.g., ensuring that binary operators receive exactly two children). The RNN outputs a probability distribution over the available tokens in the function library (\autoref{fig:dso}b), and the next token is sampled from this distribution. The process repeats until a complete expression tree (all leaf nodes filled) is generated \cite{petersen2019deep}, and the sampled token sequence is decoded into the corresponding analytical expression (\autoref{fig:dso}c).
    
    In each epoch, a batch of $N$ expressions is sampled and used to update the RNN policy (\autoref{fig:dso}d). The generation procedure for each expression consists of three phases:
    
    \begin{enumerate}
        
        \item \textbf{Initialization}: The first token (the root of the expression tree) is sampled from a predefined function library containing operators (e.g., $+$, $\times$), mathematical functions (e.g., $\sin$, $\exp$, $\log$), constants, and independent variables. At this stage no parent or sibling context is available, so the RNN receives empty inputs and samples the root token from its initial output distribution, subject to the prior constraints.
        
        \item \textbf{Sequential generation}: After each token is sampled, the partial traversal is extended and the RNN's observations are updated to reflect the new parent and sibling relationships. User-defined prior constraints can be imposed at this stage to exclude syntactically invalid or physically meaningless expressions, for example, preventing all children of a binary operator from being constants, or preventing trigonometric functions from being nested within other trigonometric functions. These priors restrict the search space without eliminating valid expressions.
        
        \item \textbf{Termination and evaluation}: The generation terminates once all terminal nodes (constants or variables) have been sampled, yielding a complete expression tree. The expression is evaluated on the training data and a scalar reward is computed, which is used to update the RNN policy through reinforcement learning.
        
    \end{enumerate}
    
    The reward function in DSO is derived from the normalized root mean square error (NRMSE), defined as
        \begin{equation}
            \mathrm{NRMSE} = \frac{1}{\sigma_y} 
            \sqrt{\frac{1}{n} \sum_{i=1}^{n} \left( y_i - f(x_i) \right)^2},
        \end{equation}

    \noindent where $(x_i, y_i)$ for $i = 1, \dots, n$ denotes the $n$ training data points, $f(\cdot)$ denotes the candidate expression, and $\sigma_y$ represents the standard deviation of the target values. A reward squashing function maps NRMSE to a bounded reward \cite{petersen2019deep} as
        \begin{equation}
            R({\tau}) = \frac{1}{1 + \mathrm{NRMSE}},
        \end{equation}

    \noindent so that $R \to 1$ for a perfect fit and $R \to 0$ for a very poor one. \\
        
    Crucially, the policy is optimized using a \emph{risk-seeking} policy gradient \cite{petersen2019deep}. Whereas the standard objective maximizes the \emph{expected} (mean) reward over all sampled expressions, the risk-seeking objective maximizes the expected reward of only the top-$\varepsilon$ fraction of generated expressions. Formally, for a batch of $N$ expressions the gradient is estimated as
    \begin{equation}
        \begin{split}
            \nabla_{\theta} J_{\mathrm{risk}}(\theta; \varepsilon) &\approx \frac{1}{\varepsilon N} \sum_{i=1}^{N}
            \left( R(\tau^{(i)}) - \tilde{R}_{\varepsilon} \right) \\
            &\quad \times\, \mathbf{1}_{R(\tau^{(i)}) \ge \tilde{R}_{\varepsilon}}\,
            \nabla_{\theta} \log p(\tau^{(i)} \mid \theta),
        \end{split}
    \end{equation}

    \noindent where $\tilde{R}_{\varepsilon}$ is the empirical $(1-\varepsilon)$-quantile of the batch rewards and $\mathbf{1}_{\{\cdot\}}$ denotes the indicator function. Equivalently, the sum runs only over the filtered batch $\Gamma = \{\tau^{(i)} : R(\tau^{(i)}) \ge \tilde{R}_{\varepsilon}\}$, so that only the best-scoring samples contribute to each update (\autoref{fig:dso}d). This biases the search toward exceptional expressions rather than average ones, which is appropriate when the goal is the single best analytical formula.
    
    The DSO framework also incorporates a constant optimization step: for each sampled expression, any numerical constants are fitted to the training data by maximizing the reward using a gradient-based optimizer (L-BFGS-B in this work) \cite{petersen2019deep}. This is performed as part of the reward computation, before each training step of the RNN. This allows the structure search to remain in discrete symbolic space while numerical constants are handled by continuous optimization.
    
    The DSO package additionally supports a hybrid operating mode (\texttt{gp\_meld}), introduced by Mundhenk et al. \cite{mundhenk2021}, which couples neural-guided search with an inner genetic programming loop. In this mode, the most recent batch of expressions sampled by the RNN serves directly as the starting population for $S$ generations of GP. The GP step applies the crossover and mutation operations described in \autoref{sec:SR} to produce refined offspring; the best-performing GP samples are then included in the policy gradient update for the RNN. This process repeats for each new RNN batch, so that the GP starting population improves as the RNN learns. The hybrid approach combines the stateful learning of the RNN, which accumulates knowledge across batches through its parameters $\theta$, with the large structural variation introduced by GP. This combination produces expressions that are essentially out-of-distribution from the current RNN policy and helps escape local optima \cite{mundhenk2021}.
 
\subsection{PyOperon Framework for Symbolic Regression}
\label{sec:PyOperon}
    
\begin{figure}
        \centering
        \includegraphics[width=\columnwidth]{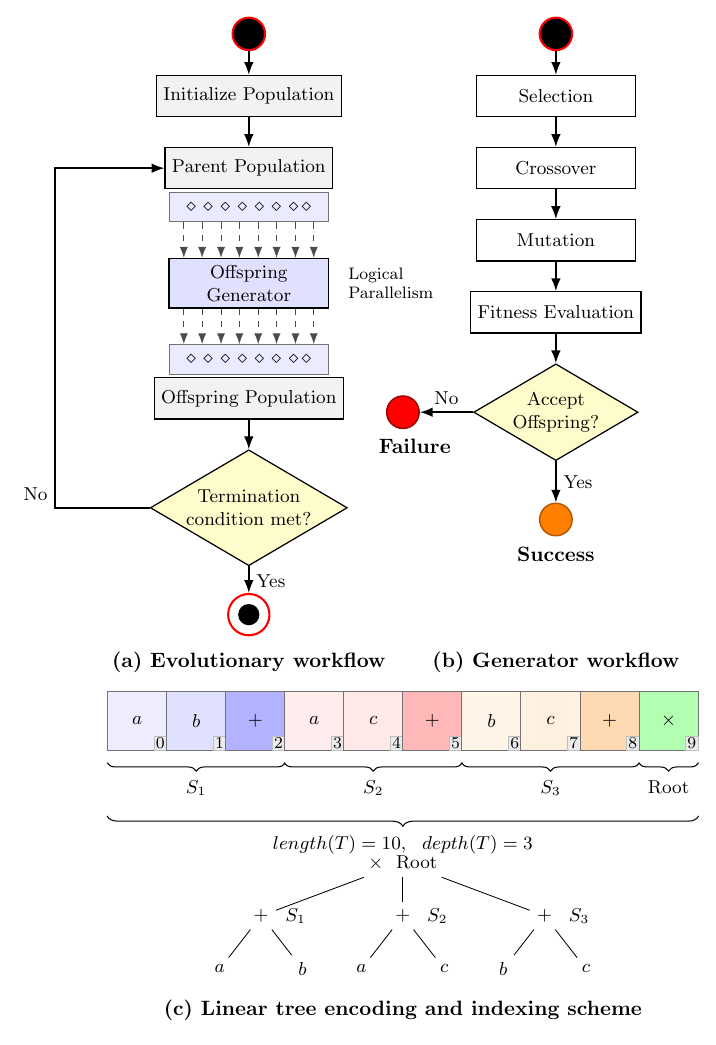}
        \caption{Overview of the PyOperon symbolic regression framework.
        \textbf{(a)} The generational evolutionary loop: an initial population is created, and in every generation the offspring generator is queried repeatedly, with each query executed as an independent logical thread, until a new population is filled or a termination criterion is met.
        \textbf{(b)} The workflow of a single offspring generator: parent selection, crossover, mutation, and fitness evaluation produce a recombinant that is either accepted (\emph{success}) or rejected (\emph{failure}) according to a configurable acceptance criterion.
        \textbf{(c)} The compact linear (postfix) tree encoding: an expression tree is stored as a contiguous array of nodes indexed in post-order, so that every subtree ($S_1, S_2, S_3$) occupies a contiguous block and the root is the last element. Adapted from Burlacu et al. \cite{burlacu2023symbolic}.}
        \label{fig:operon}
    \end{figure}

    In addition to DSO, this work employs \texttt{PyOperon} \cite{pyoperon}, the Python bindings of Operon, a modern C++ framework for symbolic regression \cite{burlacu2023symbolic}. In contrast to DSO, which relies primarily on an RNN policy, Operon is a genetic programming framework: it maintains a population of candidate expressions and improves them through selection and evolutionary operators across successive generations. Its overall evolutionary loop and internal structure are summarized in \autoref{fig:operon}. In the following, we refer to the framework as PyOperon.
    
    \noindent\textbf{Solution representation.}
    Candidate expressions in PyOperon are represented as expression trees stored in a compact \emph{linear} (postfix) form: each tree is a contiguous array of nodes laid out in post-order, so that child nodes always precede their parents and the root is the final element (\autoref{fig:operon}c). Since every node is a fixed-size (40-byte) plain data type element storing its symbol, arity, length, depth, and value, whole populations occupy very little memory and evaluation benefits from strong cache locality \cite{burlacu2023symbolic}. Trees are initialized by a balanced tree creator that produces controllable distributions of tree sizes and symbol frequencies, giving a well-mixed initial population. The evolutionary operators, namely subtree crossover, subtree, point, and hoist mutations, follow classical genetic programming principles \cite{Koza1992, la2021contemporary} and exploit the linear encoding to manipulate subtrees as simple array ranges.
    
    \noindent\textbf{Evaluation and constant optimization.}
    Each tree is evaluated over the dataset by an interpreter that uses vectorized arithmetic (via the Eigen library) and can run in single- or double-precision mode. Crucially, the same evaluation routine is generalized through C++ templates to operate on \emph{dual numbers} as well as scalars, which yields exact first-order derivatives by automatic differentiation \cite{burlacu2023symbolic}. During fitness evaluation, this gradient information drives a Levenberg--Marquardt nonlinear least-squares optimization of the numerical constants of each candidate expression. This refines the coefficients within a fixed structure so that fitting accuracy improves without increasing expression complexity. The optimized constants can be written back into the individual (Lamarckian model) or used only for fitness evaluation (Baldwinian model); in this work, the Lamarckian model is used. Both this local search and every fitness evaluation draw from a single global \emph{evaluation budget}, which makes different algorithmic variants directly and fairly comparable.
    
    \noindent\textbf{Evolutionary loop and offspring generation.} 
    PyOperon organizes evolution around an \emph{offspring generator}: the main generational loop (\autoref{fig:operon}a) repeatedly queries for new individuals until the offspring population is filled or termination is triggered. A single generator invocation (\autoref{fig:operon}b) selects parents, applies crossover and mutation, evaluates the resulting recombinant, and returns either a \emph{success} or a \emph{failure} according to a configurable acceptance criterion. This abstraction yields several interchangeable algorithmic variants sharing the same main loop: a \emph{basic} generator (standard GP, offspring is always accepted), a \emph{brood} generator (the tournament winner among several offsprings from the same parent pair is kept), and an \emph{offspring-selection} generator (a child is accepted only if it outperforms both parents) \cite{burlacu2023symbolic}.
    
    \noindent\textbf{Distinction from DSO and role in this work.} 
    Two features in particular distinguish PyOperon from DSO. First, the optimized constants are written back into the individuals and inherited by subsequent generations (Lamarckian learning), whereas DSO re-optimizes the constants of each sampled expression independently. Second, PyOperon supports multi-objective optimization that simultaneously minimizes prediction error and expression complexity, producing a Pareto front of candidate expressions \cite{burlacu2023symbolic}. This exposes the accuracy--simplicity trade-off directly and allows selection of expressions that are both sufficiently accurate and analytically tractable for downstream symbolic manipulation. This is a critical requirement for the FJC application in this work. Operon has demonstrated competitive performance on standard symbolic regression benchmarks \cite{la2021contemporary} and has been applied to problems ranging from the discovery of interatomic potentials in materials science to tasks requiring both high accuracy and interpretability \cite{burlacu2023symbolic}.

\section{Symbolic Regression for Fundamental Approximation Problems}
\label{sec:SRFundamentalApproximationProblems}
    
    Here, we apply the DSO and PyOperon frameworks to three approximation problems in statistical mechanics. Each subsection follows a common structure: we first introduce the physical or mathematical background of the problem and review existing analytical approximations together with their known limitations. Then, we present the approximation derived by DSO, followed by the expression obtained with PyOperon, together with the corresponding data and search settings, and compare their accuracy with the existing approximations. The three problems are ordered in the following way: the WLC force--extension relation (\autoref{WLC}); the factorial function approximation (\autoref{Factorial}); and the non-Gaussian radial probability density function of the FJC model (\autoref{Probability_function}).

\subsection{Worm-Like Chain Force–Extension Approximation}
\label{WLC}

\begin{figure}
        \centering
        \includegraphics[width=\columnwidth]{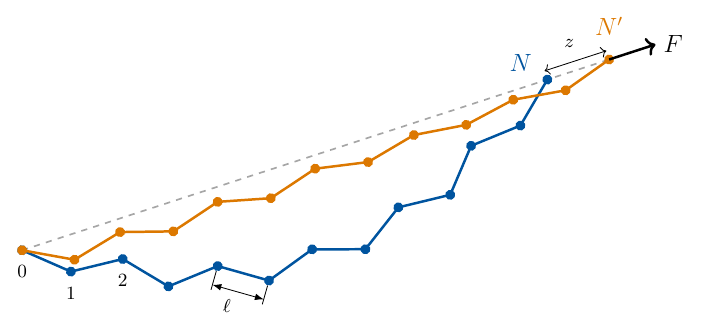}
        \caption{WLC illustration of a semi-flexible polymer chain.}
        \label{fig:wlc}
    \end{figure}

    The WLC model illustrated in \autoref{fig:wlc} describes semi-flexible polymer chains such as DNA and actin filaments as continuous elastic rods that resist bending but are inextensible and free of twist or torsion.

    In contrast to the FJC model, which assumes arbitrary rotations of chain segments with respect to each other, the WLC model accounts for the persistence of local orientation along the chain length and the angle between each neighboring nodes does not change significantly. The normalized force–extension relation for the WLC model was obtained numerically from the exact solution of the WLC Hamiltonian, and then compared with experimental data from DNA stretching experiments \cite{Bouchiat1999, Bustamante1994, Marko1995}.

    A widely used approximation formula, proposed by Marko and Siggia \cite{Marko1995}, represents the normalized force as
    \begin{equation}
        f_{\mathrm{MS}}(x) = x - 0.25 + \frac{0.25}{(1 - x)^2},
        \label{eq:marko-siggia}
    \end{equation}
    
    \noindent in terms of the relative extension $x = z/L$, where $z$ is the change in the end-to-end distance and $L=N \ell$ denotes the contour length. The normalized force is defined by  $f = Fl_p/k_BT$, where $F$ represents the applied force, $l_p$ is the persistence length, $k_B$ is the Boltzmann constant, and $T$ stands for the absolute temperature. Although Eq.~\eqref{eq:marko-siggia} provides a simple analytical form, it exhibits relative errors up to $17\%$ when compared to numerically exact WLC data.

    An improved approximation was later introduced by Ogden et al. \cite{Ogden2007}, with a quadratic correction term to the Marko–Siggia approximation as
    \begin{equation}
        f_{\mathrm{OSS}}(x)=f_{\mathrm{MS}}(x)-\frac{3}{4}x^{2} .
        \label{eq:wlc_oss}
    \end{equation}
    
    This modification significantly reduces relative error to less than $1.9\%$ across the full extension range. Further, Petrosyan~\cite{Petrosyan2017} proposed the expression:
    \begin{equation}
        f_{\mathrm{P}}(x)=f_{\mathrm{MS}}(x)-0.8x^{2.15},
        \label{eq:wlc_petrosyan}
    \end{equation}
    
    \noindent reducing the relative error to \(0.9\%\) \cite{Petrosyan2017}.

\subsubsection{DSO-Derived Approximation of the WLC Force--Extension Relation}

    The DSO framework was employed to obtain an analytical approximation for the normalized force–extension relationship of the WLC model. The algorithm was trained on the numerical data of the exact WLC model of Bouchiat \textit{et al.} \cite{Bouchiat1999}. The following approximation was obtained:
    \begin{equation}
    \begin{split}
        f_{\mathrm{DSO}}(x) &= 1.008 \Big( x - 0.8\,x^{2.15} + \frac{0.25}{(1-x)^{2}} - 0.25 \Big) \\
        &\quad \times \sin\Big( 0.622\, \exp\Big[ \\
        &\qquad \tanh\Big( \Big(\frac{0.25}{(1-x)^{2}}\Big)^{2} + 1.374^{2} \Big) \Big] \Big).
    \end{split}
    \label{eq:DSO_WLC}
    \end{equation}

\begin{figure}
        \centering
        \includegraphics[width=\columnwidth]{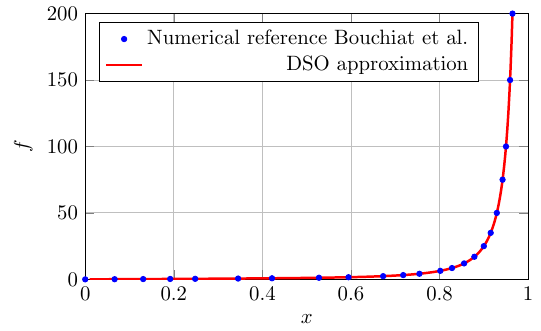}
        \caption{Numerical WLC force--extension reference values from Bouchiat et al.~\cite{Bouchiat1999} and the DSO approximation.}
        \label{fig:Test}
    \end{figure}

    The model maintains the correct asymptotic behavior of the WLC model: linear elasticity at small deformations and the force diverges as $x \to 1$ as shown in \autoref{fig:Test}.

\begin{figure}
        \centering
        \includegraphics[width=\columnwidth]{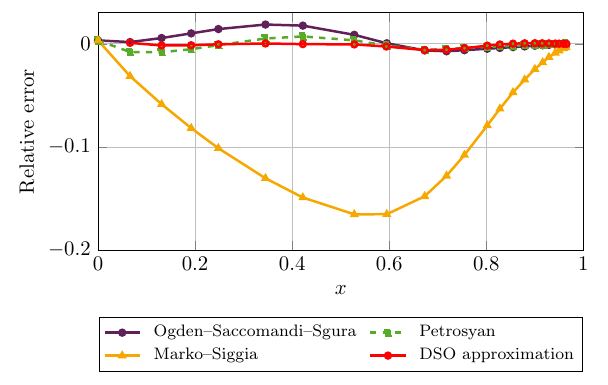}
        \caption{
            Relative error of the DSO-derived approximation.}
        \label{fig:wlc_relerr1}
    \end{figure}

    The accuracy of the DSO-derived approximation was assessed by comparison with approximations (4-6). \autoref{fig:wlc_relerr1} shows the relative error as a function of the normalized extension \(x\) across the full extension range. Accordingly, the DSO expression exhibits small and stable deviations across the full extension range, with a maximum relative error of approximately \(0.15\%\).
    
\subsubsection{PyOperon-Derived Approximation of the WLC Force--Extension Relation}
    Additionally, we obtained an analytical approximation of the worm-like chain force–extension model $f(x)$ using PyOperon (v0.6.1).
    
    Equations were limited to the length of 20 and depth of 15. The genetic algorithm used 3000 generations with a population size of 2000 through subtree crossover ($p_\mathrm{crossover}=1.0$) and multi-mutation ($p_\mathrm{mutation}=0.25$, equal weights over one-point, change-variable, change-function, replace-subtree, insert-subtree and remove-subtree operators), selected by tournament ($k=5$) with elitist reinsertion; numeric constants of every candidate were refined by a Levenberg–Marquardt step (max 10 iterations, Lamarckian). Fitness evaluation was based on $R^{2}$ of $\log f$ with linear post-fit scaling. The obtained approximation is:
    \begin{equation}
    \begin{split}
        f_{\mathrm{PyO}}(x) &= \exp\Big( \big[(0.998x)^{79.064x} - 2.903 + 6.031x\big] \\
        &\quad \times \big[\big((0.9294x)^{21.553x} + 1.143x\big)^{2.894x} \\
        &\qquad + 0.2981 - 0.2079x\big] \Big)
    \end{split}
    \label{eq:Operon_WLC}
    \end{equation}    
    
\begin{figure}
        \centering
        \includegraphics[width=\columnwidth]{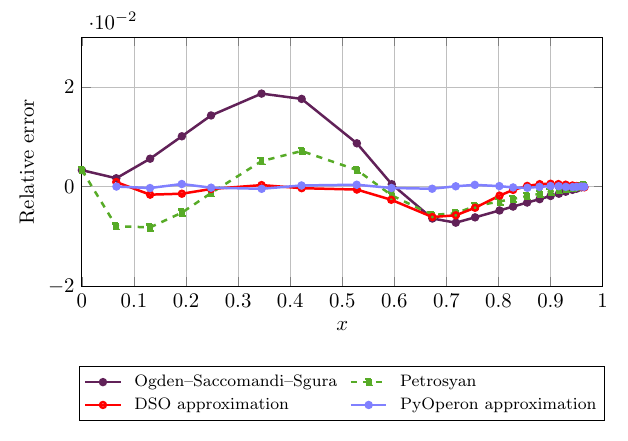}
        \caption{Comparison of relative errors for the DSO approximation, PyOperon approximation, Ogden--Saccomandi--Sgura model, and Petrosyan.}
        \label{fig:wlc_dso_operon_relerr}
    \end{figure}

    The obtained expression was compared against the Ogden–Saccomandi–Sgura and Petrosyan approximations. This expression reproduces $f(x)$ with a mean relative error of $0.02\%$ and a maximum relative error of $0.05\%$, outperforming all previously mentioned approximations including the DSO result. In \autoref{fig:wlc_dso_operon_relerr}, relative errors of these approximations are compared once more at a considerably smaller scale.

\subsection{Approximation of the Factorial Function}
\label{Factorial}
    
    The factorial $n! = \prod_{k=1}^{n} k$ is a fundamental quantity in combinatorics, probability theory, statistical mechanics, and information theory. The Gamma function $\Gamma(n+1) = n!$ represents its continuous extension and is defined for all positive real arguments. However, both $n!$ and $\Gamma(n+1)$ grow super-exponentially, which creates severe numerical instability for direct evaluation at large $n$. Moreover, they are inconvenient for analytical manipulation in their exact forms, making closed-form approximations essential from the theoretical point of view.

    The classical Stirling approximation \cite{morris2022tweaking} provides a widely used asymptotic formula:
    \begin{equation}
        n! \approx \sqrt{2\pi n} \left( \frac{n}{e} \right)^{n},
        \label{eq:stirling}
    \end{equation}
    
    \noindent or equivalently in the logarithmic form:
    \begin{equation}
        \ln(n!) \approx n \ln(n) - n + \frac{1}{2} \ln(2\pi n),
        \label{eq:log_stirling}
    \end{equation}
    
    \noindent which is more numerically stable. Stirling's formula is asymptotically exact in the sense that the relative error tends to zero as $n \to \infty$. For $n > 10$, the relative error of \eqref{eq:stirling} falls below $1\%$ and decreases monotonically with increasing $n$ \cite{robbins1955remark}. However, for small and moderate $n$ (particularly $n < 10$), the approximation error is non-negligible, limiting the utility of this formula, especially in applications involving short chains in polymer statistics \cite{Darabi2023, Morovati2019, Rubinstein2003}, the statistical mechanics of small systems \cite{bertoldi2011exact, miranda2013thermostatistics}, and information-theoretic bounds \cite{nalewajski2011elements}.
    
    Over the past century, numerous improved asymptotic approximations were proposed, see \autoref{tab:factorial_approximations}. Despite this rich history, a universal limitation persists: all existing expressions are asymptotically motivated and therefore optimized for accuracy at large $n$. Even the best analytically derived formulas, such as those of Chen \cite{chen2016more} and Hirschhorn \& Villarino \cite{Hirschhorn2013}, were designed to minimize the asymptotic error rate rather than the absolute error at small $n$. This limitation is relevant for polymer physics, where the number of chain segments $N$ may be small ($N = 2$ to $20$ in short polymer models \cite{Darabi2023, Morovati2019}), so that combinatorial descriptions of chain conformations involve factorials of small arguments. This motivates improved factorial approximations with uniformly small error specifically in the range $n \in [1, 30]$.
    
    In this work, we focus on this practically important range of $n$ and apply both DSO and PyOperon under two learning formulations: direct discovery of $\ln(n!)$ from data, and residual learning of the tail of a known continued-fraction expansion of the Stirling remainder \cite{bhatnagar2023telescoping}. Under the residual formulation, both frameworks outperform all classical approximations in \autoref{tab:factorial_approximations}, with maximum relative errors of $5.8\times10^{-10}$ (PyOperon) and $1.5\times10^{-7}$ (DSO) for $n \in [1, 30]$.
    
    \begin{table}
        \caption{Approximations of the factorial function from the literature}
        \label{tab:factorial_approximations}
        \centering
        \renewcommand{\arraystretch}{1.7}
        \setlength{\tabcolsep}{7pt}
        \small

        \begin{tabular}{p{2.3cm}p{5.2cm}}
        \toprule
        \textbf{Author} & \textbf{Approximation of $n!$} \\
        \midrule

        P.-S.\ Laplace \cite{laplace1820theorie, morris2022tweaking} (1820)
            & $\approx e^{-n} n^{n+1/2}\sqrt{2\pi}
              \!\left(1 + \dfrac{1}{12n} + \dfrac{1}{288n^2}
              + \dots\right)$ \\

        W.\ Burnside \cite{burnside1917rapidly, morris2022tweaking} (1917)
            & $\approx \sqrt{2\pi}\left(\dfrac{n + 1/2}{e}
              \right)^{n + 1/2}$ \\
    
        R.\ W.\ Gosper \cite{gosper1978decision, morris2022tweaking} (1978)
            & $\approx \sqrt{\pi}\left(\dfrac{n}{e}\right)^n
               \sqrt{2n + \tfrac{1}{3}}$ \\

        S.\ Ramanujan \cite{ramanujan1988lost} (1988)
            & $\approx \sqrt{\pi}\left(\dfrac{n}{e}\right)^n
              \!\left(8n^3 + 4n^2 + n
              + \tfrac{\theta_n}{30}\right)^{1/6}$ \\
            & {\footnotesize where $\theta_n \to 1$ as
              $n \to \infty$ and $\tfrac{3}{10} < \theta_n < 1$} \\
    
        R.\ H.\ Windschitl \cite{wang2019quicker} (2002)
            & $\approx \sqrt{2\pi n}\left(\dfrac{n}{e}\right)^n
              \left(n\sinh\!\left(\tfrac{1}{n}\right)\right)^{n/2}$ \\
    
        Gergö Nemes \cite{Nemes2010} (2010)
            & $\approx \sqrt{2\pi n}\left(\dfrac{n}{e}\right)^n
              \left(1 + \dfrac{1}{12n^2 - 1/10}\right)^n$ \\
    
        Cristinel Mortici \cite{mortici2011substantial} (2011)
            & $\approx \sqrt{2\pi n}\left(\dfrac{n}{e}
              + \dfrac{1}{12en}\right)^n$ \\
    
        M.\ D.\ Hirschhorn \&\ M.\ B.\ Villarino \cite{Hirschhorn2013} (2013)
            & Refined Ramanujan form with correction
              $\theta_n = 1 - \dfrac{11}{8n}
              + \dfrac{79}{112n^2}$ \\
              
        Chao-ping Chen \cite{chen2016more} (2016) & $\approx \sqrt{2\pi n}\left(\frac{n}{e}\right)^n \left(1     + \frac{1}{12n^3 + 24n/7 - 1/2}\right)^{n^2+\frac{53}{210}}$     \\
    
        \bottomrule
        \end{tabular}
    \end{table}
    
\subsubsection{Direct Discovery of \texorpdfstring{$\ln(n!)$}{ln(n!)} via Symbolic Regression}
\label{sec:dso_direct}
 
    Training data for the target function $f(n) = \ln(n!) = \ln\Gamma(n+1)$ were generated with the step size of $0.01$ over four domain ranges: $n \in [0,10]$, $[0,20]$, $[0,30]$, and $[0,100]$. An 80:20 training--test split was applied in all cases. Both DSO and PyOperon were trained to discover the full function $\ln(n!)$ only on the basis of these data and with no prior assumption of Stirling structure. Both frameworks used the same function library $\{+, -, \times, \div, \ln, \text{const}\}$ and a maximum expression length of 25. DSO was run in the hybrid \texttt{gp\_meld} mode (100 GP generations per batch, crossover and mutation probabilities of 0.5, tournament size 5) with $1.5\times10^{4}$ sampled expressions. PyOperon was run for 1000 generations with a population size of 1000, maximum tree lengths of 15 and 25, and three random seeds, and the candidate with the smallest maximum relative error on $n \in [1, 30]$ was selected.
    
    \paragraph{DSO results.}
    The dominant expression structure discovered by DSO across all training ranges is:
    \begin{equation}
        \ln(n!) \;=\; n\,\ln\!\left(0.36\,n + 0.36\,\ln(n + C) - \varepsilon\right),
        \label{eq:DSO_factorial}
    \end{equation}
    
    \noindent where $C$ and $\varepsilon$ are constants that vary with the training domain (\autoref{tab:dso_r2_results}).
    
    \begin{table}
        \caption{DSO-discovered expressions for $\ln(n!)$ across different training ranges.}
        \label{tab:dso_r2_results}
        \centering
        \footnotesize
        \renewcommand{\arraystretch}{1.6}
        \setlength{\tabcolsep}{7pt}
        \begin{tabular}{l p{5.5cm}}
        \toprule
        \textbf{Training range} & \textbf{DSO expression} \\
        \midrule
        $n \in [0,10]$  &
            $n\cdot\ln\!\left(0.36n + 0.36\cdot\ln(n+5.017) - 0.009\right)$ \\
        $n \in [0,20]$  &
            $n\cdot\ln\!\left(0.36n + 0.36\cdot\ln(n+6.889) - 0.092\right)$ \\
        $n \in [0,30]$  &
            $n\cdot\ln\!\left(0.36n + 0.36\cdot\ln(n+8.412) - 0.10\right)$  \\
        $n \in [0,100]$ &
            $(n+0.415)\cdot\ln\!\left(0.368n + 0.589\right)$ \\
        \bottomrule
        \end{tabular}
    \end{table}
    
    It can be seen that the discovered structure closely mirrors Stirling's dominant term $n\ln n$, where the argument of the logarithm is approximately linear in $n$, yet DSO arrived at this form autonomously without any structural prior. Furthermore, across the three smaller ranges ($n \leq 30$), only the shift constant $C$ and the offset $\varepsilon$ vary, while the coefficients $0.36$ remains stable. This indicates that DSO has identified a genuine structural property of $\ln(n!)$ rather than overfitting to a specific domain. The shift $C$ increases with the training range, compensating for the growing asymptotic regime included in the data. For $n \in [0,100]$, the discovered expression takes a simpler linear-logarithmic form, which is the natural asymptotic structure at large $n$.
    
    \paragraph{PyOperon results.}
    For the training range $n \in [0, 30]$, PyOperon discovers
    \begin{equation}
    \begin{split}
        \ln(n!) \;\approx\;& (0.999697\,n + 0.492776) \\
        &\times \ln(1.180865\,n + 1.001234) - 1.164677\,n .
    \end{split}
        \label{eq:operon_direct}
    \end{equation}
 
    \noindent Since $\ln(1.180865\,n) = \ln n + 0.1662$, the expression is approximately $(n + \tfrac{1}{2})\ln n - n$ for large $n$, i.e., PyOperon recovers the leading terms of the logarithmic Stirling formula \eqref{eq:log_stirling}, again without a structural prior.
 
    Both direct expressions capture the correct global structure of $\ln(n!)$. Their maximum relative errors on $n \in [1, 30]$ are $7.5\times10^{-2}$ (DSO) and $1.2\times10^{-4}$ (PyOperon), so that the direct formulation does not improve upon the dedicated analytical approximations. This motivates the residual strategy presented next.

\subsubsection{Hybrid Stirling--Correction Formula via Residual Learning}
\label{sec:operon_cf}
    
    While the direct approach above treats $\ln(n!)$ as an unknown function to be discovered, it is also possible to exploit the known mathematical structure of the Stirling residual, defined as
    \begin{equation}
        R(n) = \ln(n!) - \left( n\ln n - n + \tfrac{1}{2}\ln(2\pi n) \right).
    \end{equation}
    
    \noindent While the classical Laplace asymptotic series provides only a divergent approximation of $R(n)$, Bhatnagar and Rajkumar \cite{bhatnagar2023telescoping} showed that $R(n)$ admits an equivalent convergent continued-fraction (CF) expansion via the method of telescoping continued fractions as
    \begin{equation}
        R(n) \;=\; \cfrac{1}{12n + \cfrac{2}{5n + \cfrac{53}{42n + w(n)}}},
        \label{eq:cf_structure}
    \end{equation}
    
    \noindent where the constants $12$, $2$, $5$, $53$, and $42$ are analytically exact values \cite{bhatnagar2023telescoping}, and only the fourth-level tail $w(n)$ is unknown. Both DSO and PyOperon were trained to discover an approximation $\hat{w}(n)$ of this tail from data, replacing the divergent asymptotic series with a compact, data-driven expression at the deepest level of the fraction. Accordingly,
    \begin{equation}
        \ln(n!) \;\approx\; n\ln n - n + \tfrac{1}{2}\ln(2\pi n)
        \;+\; \cfrac{1}{12n + \cfrac{2}{5n +
        \cfrac{53}{42n + \hat{w}(n)}}},
        \label{eq:operon_cf}
    \end{equation}

    \noindent where $\hat{w}(n)$ is the SR-discovered expression.
    
    \paragraph{Training setup.}
    The training target $w(n)$ is computed by algebraically inverting the three known CF levels as
    \begin{equation}
        w(n) \;=\; \frac{53}{\,\dfrac{2}{\,\frac{1}{R(n)} - 12n\,}- 5n\,} - 42n ,
        \label{eq:w_inversion}
    \end{equation}
    
    \noindent evaluated at 40-digit arithmetic precision on $n \in [1, 30]$ with the step size of $0.01$ to eliminate floating-point noise in the training targets. Both frameworks used the function library $\{+, -, \times, \div, \text{const}\}$ and three random seeds. PyOperon was applied across four configurations (maximum tree lengths 5, 8, 12 and MSE/R$^2$ objectives), yielding 12 candidate expressions. DSO was applied in the \texttt{gp\_meld} mode with the settings of the direct experiment, maximum lengths of 12 and 20, and $5\times10^{4}$ sampled expressions per run, yielding 6 candidate expressions. For both frameworks, the best candidate was selected by the same minimax criterion: minimizing the worst-case ratio of model error to Chen~(2016) error over a dense screening grid. Numerical constants were subsequently refined by Nelder--Mead optimization targeting the same minimax criterion.
    
    \paragraph{Discovered expression.}
    For the training range $n \in [1, 30]$, PyOperon discovers:
    \begin{equation}
        \hat{w}_{\mathrm{PyO}}(n) \;=\; 0.000124 \;+\; \frac{836.637 + 547.677\,n}{24.522\,n^2 + 37.150\,n + 37.347},
        \label{eq:w_hat}
    \end{equation}
    
    \noindent and DSO discovers:
    \begin{equation}
        \hat{w}_{\mathrm{DSO}}(n) \;=\; \cfrac{22.101}{n + \cfrac{1.2505}{n + 1.1432}}.
        \label{eq:w_hat_dso}
    \end{equation}
    
    \noindent Inserting either tail into \eqref{eq:operon_cf} yields the complete hybrid formula. Both expressions are rational functions of $n$, which are differentiable and integrable in closed form at every level of the continued fraction, making them directly applicable to analytical calculations in polymer physics.
   
    Remarkably, all six DSO runs converged to \eqref{eq:w_hat_dso}, which is itself a two-level continued fraction. The exact continuation of \eqref{eq:cf_structure} is $w(n) = c_1/(n + c_2/(n + \dots))$ with $c_1 = 8190/371 \approx 22.0755$ and $c_2 = 22999/22737 \approx 1.0115$, so that DSO recovers the next two levels of the exact expansion without any structural prior, with a leading coefficient within $0.12\%$ of the exact value. In contrast, the constant term in \eqref{eq:w_hat} does not follow the asymptotic $1/n$ decay of $w(n)$.
   
\subsubsection{Comparison of Learning Formulations and Frameworks}
\label{sec:factorial_comparison}

    All approximations were evaluated on $n \in [1, 30]$ with the step size of $0.01$ by the relative error of $n!$,
    \begin{equation}
        \varepsilon(n) \;=\; \left| \frac{\widehat{n!}}{n!} - 1 \right|
        \;=\; \left| \exp\!\left( \widehat{\ln(n!)} - \ln\Gamma(n+1) \right) - 1 \right| ,
        \label{eq:factorial_error}
    \end{equation}
    
    \noindent computed in 40-digit arithmetic, since the errors of the best approximations approach or fall below double precision for $n \gtrsim 10$. Since DSO and PyOperon were applied to identical targets in both formulations, the effect of the learning formulation can be separated from that of the search framework (\autoref{tab:factorial_strategy}).

    \begin{table}
        \caption{Maximum and mean relative error \eqref{eq:factorial_error} of $n!$ on $n \in [1,30]$ for both frameworks under both learning formulations, compared with the classical approximations.}
        \label{tab:factorial_strategy}
        \centering
        \footnotesize
        \renewcommand{\arraystretch}{1.3}
        \setlength{\tabcolsep}{4pt}
        \begin{tabular}{llcc}
            \toprule
            Formulation & Framework & max.\ $\varepsilon$ & mean $\varepsilon$ \\
            \midrule
            Direct, $\ln(n!)$   & DSO, \eqref{eq:DSO_factorial}      & $7.46\times10^{-2}$  & $3.89\times10^{-2}$ \\
                                & PyOperon, \eqref{eq:operon_direct} & $1.19\times10^{-4}$  & $3.39\times10^{-5}$ \\
            \midrule
            CF tail, $w(n)$     & DSO, \eqref{eq:w_hat_dso}          & $1.45\times10^{-7}$  & $6.03\times10^{-10}$ \\
                                & PyOperon, \eqref{eq:w_hat}         & $5.81\times10^{-10}$ & $3.05\times10^{-11}$ \\
            \midrule
            \multicolumn{2}{l}{Stirling, \eqref{eq:stirling}}                     & $7.79\times10^{-2}$ & $9.63\times10^{-3}$ \\
            \multicolumn{2}{l}{Chen (2016) \cite{chen2016more}}                     & $1.40\times10^{-4}$ & $1.03\times10^{-6}$ \\
            \multicolumn{2}{l}{Hirschhorn \& Villarino (2013) \cite{Hirschhorn2013}} & $7.52\times10^{-6}$ & $1.85\times10^{-7}$ \\
            \bottomrule
        \end{tabular}
    \end{table}

    \begin{figure}
        \centering
        \includegraphics[width=\columnwidth]{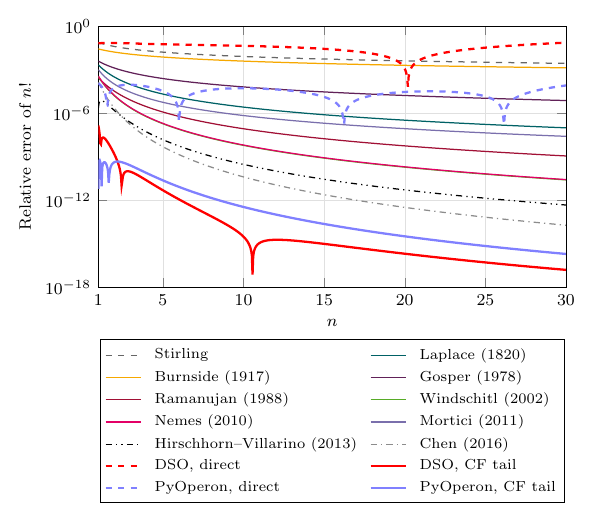}
        \caption{Relative error \eqref{eq:factorial_error} of the classical factorial approximations of \autoref{tab:factorial_approximations} and of the expressions obtained in this work by direct learning of $\ln(n!)$ and by residual learning of the continued-fraction tail $\hat{w}(n)$, for $n \in [1, 30]$.}
        \label{fig:factorial_error}
    \end{figure}
 
    The choice of formulation has the dominant effect on accuracy. Replacing direct learning by residual learning of the CF tail reduces the maximum error by a factor of approximately $5\times10^{5}$ for DSO and $2\times10^{5}$ for PyOperon. Under the residual formulation, both frameworks outperform all classical approximations (\autoref{fig:factorial_error}). The PyOperon tail \eqref{eq:w_hat} reduces the maximum error by a factor of $1.3\times10^{4}$ relative to the most accurate classical formula on this interval, that of Hirschhorn and Villarino \cite{Hirschhorn2013}, and by a factor of $2.4\times10^{5}$ relative to Chen~(2016) \cite{chen2016more}. The DSO tail \eqref{eq:w_hat_dso} improves upon these formulas by factors of approximately $50$ and $10^{3}$, respectively.
    
    The high accuracy of the residual formulation results from the position of $w(n)$ at the deepest level of the continued fraction: a deviation of $\hat{w}(n)$ from $w(n)$ enters $\ln(n!)$ only through the sensitivity $\partial R/\partial w$, which decreases from $6.3\times10^{-6}$ at $n = 1$ to $2.3\times10^{-14}$ at $n = 30$. Consequently, tails that reproduce $w(n)$ only to within about $1\%$ still yield relative errors of $n!$ below $10^{-6}$.
    
    For identical targets, PyOperon yields the smaller maximum error in both formulations, by factors of approximately $600$ (direct) and $250$ (CF tail). In the direct formulation, the maximum error of the PyOperon expression is comparable to that of Chen~(2016), but its mean error is larger by a factor of about $30$. In the residual formulation, the maximum errors of both tails occur at $n \approx 1$, whereas for $n \gtrsim 2.2$, i.e., on $96\%$ of the interval, the DSO tail is more accurate, consistent with its continued-fraction structure. The same structure leads to better extrapolation: at $n = 100$, the relative errors are $4.2\times10^{-21}$ (DSO) and $4.6\times10^{-20}$ (PyOperon), compared with $4.2\times10^{-18}$ for Chen~(2016), whereas the direct PyOperon expression \eqref{eq:operon_direct} degrades to $1.0\times10^{-2}$.
    
\subsection{Approximation of the FJC Model Radial Probability Density}
\label{Probability_function}

    Freely jointed chain (FJC) model represents a polymer chain as a sequence of $n$ rigid links of equal length $\ell$, connected by freely rotating joints. The orientations of the individual links are assumed to be statistically independent and isotropically distributed \cite{Rubinstein2003}. The end-to-end vector of a chain configuration is
    \begin{equation}
        \mathbf{R} = \sum_{i=1}^{n}\mathbf{r}_{i},
        \qquad
        \lVert\mathbf{r}_{j}\rVert=\ell,
        \quad j=1,\ldots,n,
        \label{eq:FJC_end_to_end_vector}
    \end{equation}
    
    \noindent where $\mathbf{r}_{i}$ is the vector associated with the $i$-th link. We denote the magnitude of the end-to-end vector by $R=\lVert\mathbf{R}\rVert$ and introduce the relative chain stretch
    \begin{equation}
        \lambda_r = \frac{R}{n\ell},\qquad 0\leq\lambda_r\leq1 . 
        \label{eq:FJC_normalized_extension}
    \end{equation}

    The three-dimensional spatial probability density of the end-to-end vector is given by \cite{Darabi2023}
    \begin{equation}
        P_{\mathrm{3D}}(R,n)
        =
        \frac{1}{2\pi^{2}R}
        \int_{0}^{\infty}
            \rho\sin(\rho R)
            \left[
                \frac{\sin(\rho\ell)}{\rho\ell}
            \right]^{n}
        \mathrm{d}\rho,
        \label{eq:FJC_exact_spatial_density}
    \end{equation} for \(0<R<n\ell\).

    The density satisfies the three-dimensional normalization condition as
    \begin{equation}
        4\pi
        \int_{0}^{n\ell}
            R^{2}P_{\mathrm{3D}}(R,n)
        \,\mathrm{d}R
        = 1.
        \label{eq:FJC_spatial_normalization}
    \end{equation}

    Although \eqref{eq:FJC_exact_spatial_density} can be evaluated numerically, its oscillatory integral form is inconvenient for repeated evaluation and subsequent analytical operations. For the subsequent comparison, it is convenient to work with the dimensionless radial probability density in terms of $\lambda_r$. For an isotropic three-dimensional distribution, the radial density with respect to \(R\) is obtained by multiplying the spatial density by the spherical Jacobian \(4\pi R^{2}\).
    \begin{align}
        p_{\mathrm{exact}}(\lambda_r,n)
        &=
        4\pi R^{2}
        P_{\mathrm{3D}}(R,n)
        \frac{\mathrm{d}R}{\mathrm{d}\lambda_r}
        \notag
        \\
        &=
        4\pi n^{3} \ell^{3}\lambda_r^{2}
        P_{\mathrm{3D}}(n\ell \lambda_r,n).
        \label{eq:FJC_exact_radial_density}
    \end{align}
    
    It is normalized according to
    \begin{equation}
        \int_{0}^{1}
            p_{\mathrm{exact}}(\lambda_r,n)
        \,\mathrm{d}\lambda_r
        = 1.
        \label{eq:FJC_radial_normalization}
    \end{equation}
    
    A widely used non-Gaussian approximation of FJC is the classical Kuhn--Gr\"un inverse-Langevin approximation \cite{Kuhn1942}. Introducing
    \begin{equation}
        \beta(\lambda_r)
        =
        \mathcal{L}^{-1}(\lambda_r),
        \qquad
        \mathcal{L}(x)
        =
        \coth x-\frac{1}{x},
        \label{eq:FJC_inverse_Langevin}
    \end{equation}

    \noindent where \(\mathcal{L}\) denotes the Langevin function, the conventional Kuhn--Gr\"un spatial     approximation can be expressed, up to factors independent of \(\lambda_r\) at fixed \(n\), through the     inverse-Langevin term
    \[
        \left[
            \frac{\sinh\beta(\lambda_r)}{\beta(\lambda_r)}
            \exp\!\bigl(-\lambda_r\beta(\lambda_r)\bigr)
        \right]^n.
    \]
    Transforming the conventional Kuhn--Gr\"un spatial approximation \cite{Darabi2023} to a radial density with     respect to \(\lambda_r\), and omitting factors independent of \(\lambda_r\) at fixed \(n\), we obtain
    \begin{equation}
        q_{\mathrm{KG}}(\lambda_r,n)
        =
        \lambda_r^{2}
        \left[
            \frac{\sinh\beta(\lambda_r)}{\beta(\lambda_r)}
            \exp\!\bigl(-\lambda_r\beta(\lambda_r)\bigr)
        \right]^{n}.
        \label{eq:FJC_KG_unnormalized}
    \end{equation}

    The Kuhn--Gr\"{u}n approximation takes into account finite chain extensibility and approaches the Gaussian-chain result in the long-chain, small-extension limit. Its accuracy, however, deteriorates for short chains and at large extensions \cite{Darabi2023, Morovati2019}. Darabi et al. \cite{Darabi2023} proposed a Pad\'{e}-type correction to the Kuhn--Gr\"{u}n approximation as follows
    \begin{equation}
        q_{\mathrm{D}}(\lambda_r,n)
        =
        q_{\mathrm{KG}}(\lambda_r,n)
        K_{\mathrm{D}}(\lambda_r,n),
        \label{eq:FJC_Darabi_unnormalized}
    \end{equation}
    where
    \begin{equation}
        K_{\mathrm{D}}(\lambda_r,n)
        =
        \frac{
            a_{\mathrm{D}}(n)\lambda_r^{3}
            +
            b_{\mathrm{D}}(n)\lambda_r^{2}
            +
            c_{\mathrm{D}}(n)\lambda_r
            +
            1
        }{
            (1-\lambda_r)^{2}
        }.
        \label{eq:FJC_Darabi_correction}
    \end{equation}

    The reported coefficient functions are
    \begin{align}
        a_{\mathrm{D}}(n)
        &=
        -0.42412
        +
        \frac{0.44}{n},
        \label{eq:FJC_Darabi_a}
        \\
        b_{\mathrm{D}}(n)
        &=
        1.2675
        +
        \frac{
            \ln(25.8579\,n)
        }{
            36.8362
        },
        \label{eq:FJC_Darabi_b}
        \\
        c_{\mathrm{D}}(n)
        &=
        -1.8427
        -
        \frac{n}{4018.4835}.
        \label{eq:FJC_Darabi_c}
    \end{align}
    
    For each chain length \(n\), Darabi et al.\ \cite{Darabi2023} determined the coefficients \(a\), \(b\), and \(c\) of the Padé correction by Levenberg--Marquardt optimization, minimizing the squared error in \(\ln P\). The resulting discrete coefficient values were then fitted as functions of \(n\), yielding the expressions \eqref{eq:FJC_Darabi_a}--\eqref{eq:FJC_Darabi_c}.
    
\subsubsection{Reference Data and Numerical Evaluation}
    \paragraph{Reference data.}
    The reference dataset was obtained using Maple by evaluating the exact Fourier representation \cite{Darabi2023} in \eqref{eq:FJC_exact_spatial_density} with \(\ell=1\) for twenty chain lengths \(n\in\{5,10,\ldots,100\}\) and 100 unique values of the relative chain stretch \(\lambda_r\in\{0.01,0.02,\ldots,1.00\}\).

    For each pair \((\lambda_r,n)\), the Maple worksheet first evaluates \(P_{\mathrm{3D}}(n\lambda_r,n)\) and stores the corresponding radial density with respect to the end-to-end distance R,
    \begin{equation}
        p_R(R,n) = 4\pi R^{2}P_{\mathrm{3D}}(R,n), \qquad R=n\lambda_r.
        \label{eq:FJC_maple_radial_density}
    \end{equation}

    \noindent The dimensionless density with respect to \(\lambda_r\) is obtained in the Python post-processing by normalizing each fixed-\(n\) curve separately as,
    \begin{equation}
        p_{\mathrm{exact}}(\lambda_r,n)
        =
        \frac{
            p_R(n\lambda_r,n)
        }{
            \displaystyle
            \int_{0}^{1}
            p_R(ns,n)\,\mathrm{d}s
        }
        \label{eq:FJC_reference_normalization}.
    \end{equation}

    \paragraph{Error measures.}
    For a normalized approximation \(\widehat{p}(\lambda_r,n)\), the logarithmic and relative density errors are defined as
    \begin{align}
        \varepsilon_{\log p}(\lambda_r,n)
        &=
        \left|
            \ln
            \frac{
                \widehat{p}(\lambda_r,n)
            }{
                p_{\mathrm{exact}}(\lambda_r,n)
            }
        \right|,
        \label{eq:fjc_err_log}
        \\
        \varepsilon^{\mathrm{rel}}_{p}(\lambda_r,n)
        &=
        \frac{
            \left|
                \widehat{p}(\lambda_r,n)
                -
                p_{\mathrm{exact}}(\lambda_r,n)
            \right|
        }{
            p_{\mathrm{exact}}(\lambda_r,n)
        }
        \times100\% .
        \label{eq:fjc_err_rel}
    \end{align}

    The mean and global maximum of $\varepsilon_{\log p}$ are denoted by $\overline{\varepsilon}_{\log p}$ and $\varepsilon_{\log p}^{\max}$, respectively. At fixed $n$, we define
    \begin{equation}
        \varepsilon_{n}^{\max}= \max_{\,0.01<\lambda_r<0.95}
        \varepsilon_{\log p}(\lambda_r,n),
    \end{equation}
    
    \noindent and use its average over the twenty chain lengths,
    \begin{equation}
        \left\langle
            \varepsilon_{n}^{\max}
        \right\rangle = \frac{1}{20}
        \sum_{n\in\{5,10,\ldots,100\}}
            \varepsilon_{n}^{\max},
        \label{eq:fjc_selection_metric}
    \end{equation}
    
    \noindent as the final correction-selection criterion.
\subsubsection{Physics-Informed Residual Learning and Search Setup}
\label{sec:fjc_residual_learning}
    
    \paragraph{Residual formulations.}
    Rather than approximating the exact radial density directly, we use the Kuhn--Gr\"un approximation \eqref{eq:FJC_KG_unnormalized} as a baseline and apply symbolic regression only to its residual. Since the baseline already captures the principal inverse-Langevin dependence and finite chain extensibility, the regression focuses on the remaining finite-chain correction. Learning this correction therefore reduces the complexity of the symbolic regression task compared with reconstructing the full density approximation. Before constructing the residual, the Kuhn--Gr\"{u}n density is normalized separately for each chain length. We denote the resulting normalized density by \(p_{\mathrm{KG}}(\lambda_r,n)\).
    
    The raw logarithmic residual is then defined as
    \begin{equation}
        r_{0}(\lambda_r,n)
        =
        \ln p_{\mathrm{exact}}(\lambda_r,n)
        -
        \ln p_{\mathrm{KG}}(\lambda_r,n).
        \label{eq:fjc_raw_residual}
    \end{equation}
    
    For each chain length, the residual is centered by subtracting its arithmetic mean over the training extensions as
    \begin{equation}
        r(\lambda_r,n)
        =
        r_{0}(\lambda_r,n)
        -
        \left\langle
            r_{0}(\cdot,n)
        \right\rangle_{\mathcal{T}_{\mathrm{train}}},
        \label{eq:fjc_residual}
    \end{equation}
    
    \noindent where \(\left\langle r_{0}(\cdot,n)\right\rangle_{\mathcal{T}_{\mathrm{train}}}\)
    denotes the unweighted arithmetic mean evaluated at fixed \(n\) over the training points.
    
    Two correction strategies were considered. In the first, the Pad\'{e} structure of \eqref{eq:FJC_Darabi_correction} was retained, and symbolic regression was used to find the coefficient functions \(a(n)\), \(b(n)\), and \(c(n)\). In the second, the logarithmic correction \(r(\lambda_r,n)\) was approximated directly without prescribing a functional form for the correction.

    \paragraph{Input features and search settings.} For the free residual search the function set was identical in PyOperon and DSO and contained the following seventeen entries:

    \[n,\ n^{-1},\ n^{-2},\ \ln n,
        \quad
        \lambda_r,\ \lambda_r^{2},\ \lambda_r^{3},\ \lambda_r^{4},\ \lambda_r^{5},
    \]
    \[
        \ln(1-\lambda_r),
        \quad
        (1-\lambda_r)^{-1},
        \quad
        (1-\lambda_r)^{-2},
        \quad
        \mathcal{L}^{-1}(\lambda_r),
    \]
    together with
    \begin{equation}
        \operatorname{sinc}(\lambda_r)
        =
        \frac{\sin(\pi \lambda_r)}{\pi \lambda_r},
        \label{eq:FJC_sinc}
    \end{equation}

    \noindent and $\ln[\operatorname{sinc}(\lambda_r)]$, $\sin(\pi \lambda_r)$, $\cos(\pi \lambda_r)$.

    For the coefficient-constrained search the target equations depend on $n$ alone, and both PyOperon and DSO used the following function set:
    \[
        n,\ n^{-1},\ n^{-2},\ n^{-3},\ \ln n,\ n^{-1}\ln n,\ \sqrt{n},\ n^{-1/2}.
    \]
    
    The remaining settings are summarized in \autoref{tab:fjc_settings}. The internal objectives were used only during symbolic search; all candidates were subsequently normalized and re-evaluated using $\langle\varepsilon_{n}^{\max}\rangle$.

    \begin{table}[!tb]
        \centering
        \footnotesize
        \setlength{\tabcolsep}{3pt}
        \caption{
            Settings used in the FJC residual-learning experiments.
        }
        \label{tab:fjc_settings}
        \begin{tabular*}{\columnwidth}{@{\extracolsep{\fill}}lll@{}}
            \toprule
            & PyOperon & DSO \\
            \midrule
            Operators, $\widehat{r}$
                & $+,-,\times,\div$
                & $+,-,\times,\div$ \\
                & $\sin,\cos,\exp,\ln$
                & $\sin,\cos,\exp$ \\
            Operators, $a,b,c$
                & $+,-,\times,\div$
                & $+,-,\times,\div,\ln$ \\
            Population / batch
                & $1000$
                & $500$ \\
            Runs, $\widehat{r}\,/\,a,b,c$
                & $12\ /\ 4$ per coeff.
                & $2\ /\ 1$ per coeff. \\
            Budget per run
                & $120\ /\ 25\,\mathrm{s}$
                & $6\times10^{4}\ /\ 10^{5}$ expr. \\
            Length limit
                & $21$--$31\ /\ 13$ nodes
                & $24\ /\ 16$ tokens \\
            Search algorithm
                & NSGA-2,
                & policy gradient \\
                & $(R^{2},\ \text{length})$
                & \\
            Constant optimization
                & Levenberg--Marquardt
                & L-BFGS-B \\
                & $25$ iter., Lamarckian
                & $\mathrm{gtol}=10^{-3}$ \\
            Search objective
                & $R^{2}$
                & $1/(1+\mathrm{NRMSE})$ \\
            Final selection
                & $\langle\varepsilon_{n}^{\max}\rangle$
                & $\langle\varepsilon_{n}^{\max}\rangle$ \\
            \bottomrule
        \end{tabular*}
    \end{table}

\subsubsection{Discovered Approximations}
    \paragraph{Coefficient-constrained correction.}
    In this search we retain the Pad\'e-type correction structure and use symbolic regression only to identify the coefficients \eqref{eq:FJC_Darabi_correction}. For each chain length, the coefficients of \eqref{eq:FJC_Darabi_correction} were first optimized directly using $\varepsilon_{n}^{\max}$. Symbolic regression was then used to approximate the resulting coefficient values as functions of $n$. PyOperon obtained
    \begin{align}
        a(n) &= -0.787607 + \frac{16.2819}{n^{3}} \notag \\
             &\quad + 0.0621372\ln n + \frac{0.886675}{\sqrt{n}},
        \label{eq:fjc_abc_a} \\[2pt]
        b(n) &= 1.78110 - \frac{29.0309}{n^{3}} \notag \\
             &\quad - 0.0659284\ln n - 1.09382\,\frac{\ln n}{n},
        \label{eq:fjc_abc_b} \\[2pt]
        c(n) &= -1.76942 + \frac{14.4851}{n^{3}} \notag \\
             &\quad + 1.30635\,\frac{\ln n}{n} - \frac{1.13566}{\sqrt{n}}.
        \label{eq:fjc_abc_c}
    \end{align}

    The PyOperon and DSO coefficient models give \(\left\langle \varepsilon_{n}^{\max} \right\rangle = 7.6\times10^{-3}\) and \(7.7\times10^{-3}\), respectively.

    \paragraph{Free residual correction.}
    The best free-form correction obtained by PyOperon is
    \begin{equation}
    \begin{aligned}
        \widehat r(\lambda_r,n)
        ={}& -1.976\,\lambda_r^{2} - \frac{0.00480813}{1-\lambda_r} \\
           &- 2.121\ln[\operatorname{sinc}(\lambda_r)] + F(\lambda_r,n)\,G(\lambda_r),
        \\[4pt]
        F(\lambda_r,n)
        ={}& -\frac{30.0721}{n} - \sin\!\left(5.80564\,\lambda_r^{3}\right),
        \\[4pt]
        G(\lambda_r)
        ={}& -0.0710231\,\lambda_r^{5} - 0.0246168\ln[\operatorname{sinc}(\lambda_r)].
    \end{aligned}
    \label{eq:fjc_rhat}
    \end{equation}
    
    The corresponding probability density is reconstructed by applying the obtained correction to the Kuhn--Gr\"un baseline and normalizing the result separately at each chain length.
    
    For both the PyOperon and DSO free-residual corrections, the reported errors are evaluated using the corresponding reconstructed and normalized density \(\widehat p(\lambda_r,n)\).

\subsubsection{Comparative Accuracy and Limitations}
\label{sec:fjc_comparison}

\begin{figure}
        \centering
        \includegraphics[width=\columnwidth]
            {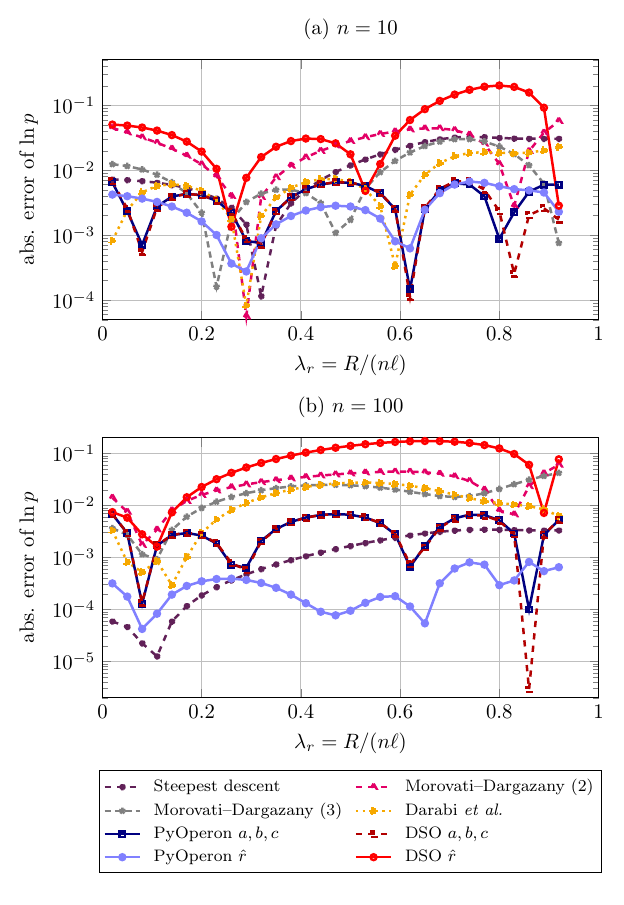}
        \caption{
            Absolute logarithmic error $\varepsilon_{\log p}$, defined in \eqref{eq:fjc_err_log}, as a function of the normalized extension for (a) $n=10$ and (b) $n=100$.
        }
        \label{fig:fjc_error_t}
    \end{figure}

\begin{figure}
        \centering
        \includegraphics[width=\columnwidth]{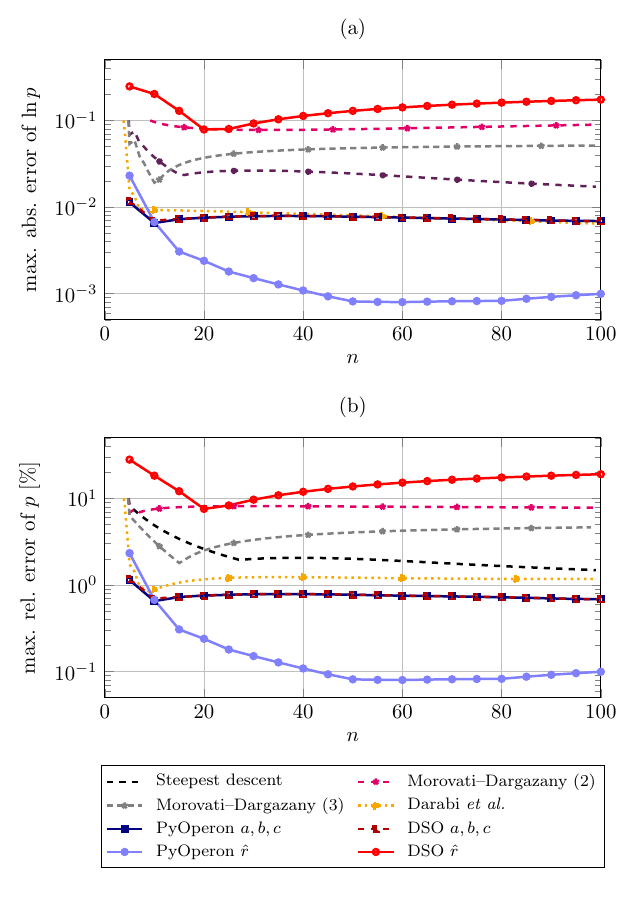}
        \caption[Maximum errors as functions of the number of links]{
            Maximum errors over \(0.01<\lambda_r<0.95\) as functions of
            the number of links \(n\): (a) maximum absolute logarithmic
            error, defined in \eqref{eq:fjc_err_log}, and
            (b) maximum relative density error, defined in
            \eqref{eq:fjc_err_rel}. All approximations are evaluated using
            the same reference data, fixed-\(n\) normalization, extension
            grid, and error definitions used in
            \autoref{tab:fjc_metrics}.
        }
        \label{fig:fjc_error_n}
    \end{figure}

    \paragraph{Comparative accuracy.}
    The errors obtained under the normalization and evaluation procedure are summarized in \autoref{tab:fjc_metrics}. Their dependence on the normalized extension and the number of links is shown in \autoref{fig:fjc_error_t} and \autoref{fig:fjc_error_n}, respectively.
    
    \begin{table*}[!tb]
        \centering
        \footnotesize
        \caption{
            Errors of the normalized radial density over
            \(0.01<\lambda_r<0.95\) for
            \(n\in\{5,10,\ldots,100\}\).
            All models were normalized separately for each chain
            length using the same trapezoidal quadrature and
            evaluated on the same tabulated grid
            \(0.01\leq\lambda_r\leq1.00\). The pointwise errors
            are defined in \eqref{eq:fjc_err_log} and
            \eqref{eq:fjc_err_rel}. Here, \(N_c\) denotes the
            number of fitted constants, and the final column gives
            the number of expression-tree nodes.
        }
        \label{tab:fjc_metrics}
        \begin{tabular*}{\textwidth}{@{\extracolsep{\fill}}lrrrrrr}
            \toprule
            Approximation
                & $\bar\varepsilon_{\log p}$
                & $\varepsilon^{\max}_{\log p}$
                & $\langle\varepsilon^{\max}_{n}\rangle$
                & $\varepsilon^{\mathrm{rel}}_{p,\max}$ [\%]
                & $N_{c}$
                & nodes \\
            \midrule
            Kuhn--Gr\"un
                & 0.7725 & 4.0223 & 3.9724 & 98.21 & -- & -- \\
            Steepest descent
                & 0.0058 & 0.0584 & 0.0117 &  6.02 & -- & -- \\
            Morovati--Dargazany (2)
                & 0.0269 & 0.0778 & 0.0755 &  7.49 & -- & -- \\
            Morovati--Dargazany (3)
                & 0.0145 & 0.0608 & 0.0389 &  6.27 & -- & -- \\
            Darabi \textit{et al.}
                & 0.0122 & 0.1498 & 0.0286 & 16.16 &  7 & -- \\
            PyOperon $a,b,c$
                & 0.0040 & 0.0113 & 0.0076 &  1.14 & 12 & 39 \\
            DSO $a,b,c$
                & 0.0040 & 0.0116 & 0.0077 &  1.16 & 13 & 46 \\
            PyOperon $\hat{r}$
                & 0.0012 & 0.0231 & 0.0026 &  2.33 &  7 & 28 \\
            DSO $\hat{r}$
                & 0.0683 & 0.2472 & 0.1433 & 28.04 &  5 & 22 \\
            \bottomrule
        \end{tabular*}
    \end{table*}

    The error metric $\langle\varepsilon_{n}^{\max}\rangle$ shows that the free PyOperon residual correction in \eqref{eq:fjc_rhat} gives the best result, $\langle\varepsilon_{n}^{\max}\rangle = 2.56\times10^{-3}$.

This error is smaller by a factor of $11.2$ than the Darabi et al. approximation defined by \eqref{eq:FJC_Darabi_correction}--\eqref{eq:FJC_Darabi_c} and by a factor of $4.6$ than the steepest-descent approximation. At $n=10$, the mean logarithmic error of the free residual is $2.97\times10^{-3}$, compared with $8.72\times10^{-3}$ for the reported coefficients \cite{Darabi2023}. At $n=100$, the corresponding values are $3.09\times10^{-4}$ and $1.36\times10^{-2}$, respectively.
    
    The free-residual PyOperon approximation achieves the lowest mean error and the smallest $\langle\varepsilon_{n}^{\max}\rangle$. In contrast, the coefficient-constrained PyOperon approximation yields lower worst-case errors, with a global maximum logarithmic error of $1.13\times10^{-2}$, compared with $2.31\times10^{-2}$ for the free-residual approximation, and a maximum relative error of $1.14\%$, compared with $2.33\%$.

    Thus, the PyOperon free residual approximation gives the best accuracy across the selected chain lengths, but it does not uniformly outperform the coefficient-constrained model. The DSO free residual approximation improves the uncorrected Kuhn--Gr\"{u}n density, but remains less accurate than the coefficient-constrained and free residual PyOperon approximations.

    \paragraph{Limitations.}
    The DSO and PyOperon searches were performed with different computational budgets and different numbers of independent runs. Their results therefore describe the expressions found under the reported configurations and should not be interpreted as a controlled comparison of the intrinsic performance of the two search algorithms.
    The highest error appears at the shortest chain, $n=5$, suggesting that an additional short-chain correction may be required for further improvement. The reported results show accuracy only on the limited extension range and do not establish generalization to intermediate values of $\lambda_r$ or to chain lengths outside $5\leq n\leq100$.

\section{Conclusion}
\label{sec:Conclusion}

    This study applied DSO and PyOperon to three approximation problems in statistical mechanics: the WLC force--extension relation, the logarithm of the factorial, and the non-Gaussian radial probability density of the FJC model. Both direct symbolic regression and residual learning strategies were investigated to obtain analytical approximations.
    
    For the WLC model, the DSO expression achieved a maximum relative error of approximately $0.15\%$, while the PyOperon expression reduced the mean and maximum errors to approximately $0.02\%$ and $0.05\%$, respectively. Compared to the best existing approximations, the maximum error was reduced by a factor of $6$ with DSO and $18$ with PyOperon.
    
    For $\ln(n!)$, both frameworks recovered the dominant structure of Stirling's formula directly from data, with maximum relative errors of $7.5\times10^{-2}$ (DSO) and $1.2\times10^{-4}$ (PyOperon) for $1 \leq n \leq 30$. Learning only the tail of the known continued-fraction expansion of the Stirling remainder \cite{bhatnagar2023telescoping} reduced these errors by more than five orders of magnitude for both frameworks. The resulting PyOperon approximation reached a maximum relative error of $5.8\times10^{-10}$, improving upon the most accurate classical formula on this interval, that of Hirschhorn and Villarino \cite{Hirschhorn2013}, by a factor of $1.3\times10^{4}$ and upon the Chen approximation \cite{chen2016more} by a factor of $2.4\times10^{5}$. The DSO approximation ($1.5\times10^{-7}$) recovered the next two levels of the exact continued fraction and was the more accurate of the two for $n \gtrsim 2$ and beyond the training range. Thus, the learning formulation had a substantially larger effect on accuracy than the choice of framework. Both hybrid approximations retain a closed form suitable for analytical differentiation.
    
    For the FJC radial probability density, the Pad\'{e}-constrained DSO and PyOperon models produced similar mean worst-case logarithmic errors, $7.7\times10^{-3}$ and $7.6\times10^{-3}$, respectively, both outperforming the Darabi et al. \cite{Darabi2023} approximation by a factor of $3.7$. The free residual PyOperon experiment reduced the error to $2.56\times10^{-3}$, corresponding to improvements by factors of $11.2$ over the Darabi et al. \cite{Darabi2023} approximation and $4.6$ over the steepest-descent approximation. The free DSO approximation was less accurate and did not improve existing models.
    
    Across all three applications, PyOperon produced the lowest numerical errors in the WLC, Stirling-residual, and free FJC experiments, whereas DSO remained competitive for the Padé-structured correction to the FJC radial probability density. Since the two frameworks were evaluated with different computational budgets and constant-optimization procedures, these results reflect the accuracy of the obtained expressions rather than the intrinsic performance of the search algorithms.

    Given its accuracy and efficiency, the proposed framework, in particular the residual-learning strategy, can be applied to other approximation problems in statistical mechanics.

\printcredits

\bibliographystyle{cas-model2-names}
\bibliography{references}

\end{document}

%% file: figs/rwth-colors.tex
\definecolor{rwth}   {RGB}{  0  84 159}
\definecolor{rwth-75}{RGB}{ 64 127 183}
\definecolor{rwth-50}{RGB}{142 186 229}
\definecolor{rwth-25}{RGB}{199 221 242}
\definecolor{rwth-10}{RGB}{232 241 250}

\definecolor{black}   {RGB}{  0   0   0}
\definecolor{black-75}{RGB}{100 101 103}
\definecolor{black-50}{RGB}{156 158 159}
\definecolor{black-25}{RGB}{207 209 210}
\definecolor{black-10}{RGB}{236 237 237}

\definecolor{magenta}   {RGB}{227   0 102}
\definecolor{magenta-75}{RGB}{233  96 136}
\definecolor{magenta-50}{RGB}{241 158 177}
\definecolor{magenta-25}{RGB}{249 210 218}
\definecolor{magenta-10}{RGB}{253 238 240}

\definecolor{yellow}   {RGB}{255 237   0}
\definecolor{yellow-75}{RGB}{255 240  85}
\definecolor{yellow-50}{RGB}{255 245 155}
\definecolor{yellow-25}{RGB}{255 250 209}
\definecolor{yellow-10}{RGB}{255 253 238}

\definecolor{petrol}   {RGB}{  0  97 101}
\definecolor{petrol-75}{RGB}{ 45 127 131}
\definecolor{petrol-50}{RGB}{125 164 167}
\definecolor{petrol-25}{RGB}{191 208 209}
\definecolor{petrol-10}{RGB}{230 236 236}

\definecolor{turkis}   {RGB}{  0 152 161}
\definecolor{turkis-75}{RGB}{  0 177 183}
\definecolor{turkis-50}{RGB}{137 204 207}
\definecolor{turkis-25}{RGB}{202 231 231}
\definecolor{turkis-10}{RGB}{235 246 246}

\definecolor{grun}   {RGB}{ 87 171  39}
\definecolor{grun-75}{RGB}{141 192  96}
\definecolor{grun-50}{RGB}{184 214 152}
\definecolor{grun-25}{RGB}{221 235 206}
\definecolor{grun-10}{RGB}{242 247 236}

\definecolor{maigrun}   {RGB}{189 205   0}
\definecolor{maigrun-75}{RGB}{208 217  92}
\definecolor{maigrun-50}{RGB}{224 230 154}
\definecolor{maigrun-25}{RGB}{240 243 208}
\definecolor{maigrun-10}{RGB}{249 250 237}

\definecolor{orange}   {RGB}{246 168   0}
\definecolor{orange-75}{RGB}{250 190  80}
\definecolor{orange-50}{RGB}{253 212 143}
\definecolor{orange-25}{RGB}{254 234 201}
\definecolor{orange-10}{RGB}{255 247 234}

\definecolor{rot}   {RGB}{204   7  30}
\definecolor{rot-75}{RGB}{216  92  65}
\definecolor{rot-50}{RGB}{230 150 121}
\definecolor{rot-25}{RGB}{243 205 187}
\definecolor{rot-10}{RGB}{250 235 227}

\definecolor{bordeaux}   {RGB}{161  16  53}
\definecolor{bordeaux-75}{RGB}{182  82  86}
\definecolor{bordeaux-50}{RGB}{205 139 135}
\definecolor{bordeaux-25}{RGB}{229 197 192}
\definecolor{bordeaux-10}{RGB}{245 232 229}

\definecolor{violett}   {RGB}{ 97  33  88}
\definecolor{violett-75}{RGB}{131  78 117}
\definecolor{violett-50}{RGB}{168 133 158}
\definecolor{violett-25}{RGB}{210 192 205}
\definecolor{violett-10}{RGB}{237 229 234}

\definecolor{lila}   {RGB}{122 111 172}
\definecolor{lila-75}{RGB}{155 145 193}
\definecolor{lila-50}{RGB}{188 181 215}
\definecolor{lila-25}{RGB}{222 218 235}
\definecolor{lila-10}{RGB}{242 240 247}